\documentclass{llncs}

\usepackage{graphicx}
\begin{document}

\title{Glottal Source Estimation using an Automatic Chirp Decomposition}

\author{Thomas Drugman $^1$, Baris Bozkurt $^2$, Thierry Dutoit $^1$}

\institute{TCTS Lab, Facult\'e Polytechnique de Mons, Belgium \and Department of Electrical \& Electronics Engineering, Izmir Institute of Technology, Turkey}
\maketitle

\begin{abstract}
In a previous work, we showed that the glottal source can be estimated from speech signals by computing the Zeros of the Z-Transform (ZZT). Decomposition was achieved by separating the roots inside (causal contribution) and outside (anticausal contribution) the unit circle. In order to guarantee a correct deconvolution, time alignment on the Glottal Closure Instants (GCIs) was shown to be essential. This paper extends the formalism of ZZT by evaluating the Z-transform on a contour possibly different from the unit circle. A method is proposed for determining automatically this contour by inspecting the root distribution. The derived Zeros of the Chirp Z-Transform (ZCZT)-based technique turns out to be much more robust to GCI location errors.
\end{abstract}

\section{Introduction}
The deconvolution of speech into its vocal tract and glottis contributions is an important topic in speech processing. Explicitly isolating both components allows to model them independently. While techniques for modeling the vocal tract are rather well-established, it is not the case for the glottal source representation. However the characterization of this latter has been shown to be advantageous in speaker recognition \cite{Speaker}, speech disorder analysis \cite{Disorder}, speech recognition \cite{Reco} or speech synthesis \cite{Synth}. These reasons justify the need of developing algorithms able to robustly and reliably estimate and parametrize the glottal signal.

Some works addressed the estimation of the glottal contribution directly from speech waveforms. Most approaches rely on a first parametric modeling of the vocal tract and then remove it by inverse filtering so as to obtain the glottal signal estimation. In \cite{DAP}, the use of the Discrete All-Pole (DAP) model is proposed. The Iterative Adaptive Inverse Filtering technique (IAIF) described in \cite{IAIF} isolates the source signal by iteratively estimating both vocal tract and source parts. In \cite{ClosedPhase}, the vocal tract is estimated by Linear Prediction (LP) analysis on the closed phase. As an extension, the Multicycle closed-phase LPC (MCLPC) method \cite{MCLPC} refines its estimation on several larynx cycles. In a fundamentally different point of view, we proposed in \cite{ZZT} a non-parametric technique based on the Zeros of the Z-Transform (ZZT). ZZT basis relies on the observation that speech is a mixed-phase signal \cite{Doval} where the anticausal component corresponds to the vocal folds open phase, and where the causal component comprises both the glottis closure and the vocal tract contributions. Basically ZZT isolates the glottal open phase contribution from the speech signal, by separating its causal and anticausal components. In \cite{Sturmel}, a comparative evaluation between LPC and ZZT-based decompositions is led, giving a significant advantage for the second technique.

This paper proposes an extension to the traditional ZZT-based decomposition technique. The new method aims at separating both causal and anticausal contributions by computing the Zeros of a Chirp Z-Transform (ZCZT). More precisely, the Z-transform is here evaluated on a contour possibly different from the unit circle. As a result, we will see that the estimation is much less sensitive to the Glottal Closure Instant (GCI) detection errors. In addition, a way to automatically determine an optimal contour is also proposed.

The paper is structured as follows. Section \ref{sec:ZZT} reminds the principle of the ZZT-based decomposition of speech. Its extension making use of a chirp analysis is proposed and discussed in Section \ref{sec:ZCZT}. In Section \ref{sec:results}, a comparative evaluation of both approaches is led on both synthetic and real speech signals. Finally we conclude in Section \ref{sec:conclu}.

\section{ZZT-based Decomposition of Speech}\label{sec:ZZT}

For a series of $N$ samples $(x(0),x(1),...x(N-1))$ taken from a
discrete signal $x(n)$, the $ZZT$ representation is defined as
the set of roots (zeros) $(Z_1,Z_2,...Z_{N-1})$ of the corresponding
Z-Transform $X(z)$:

\begin{equation}\label{eq:ZZT}
X(z)=\sum_{n=0}^{N-1} x(n)z^{-n}=x(0)z^{-N+1}\prod_{m=1}^{N-1} (z-Z_m)
\end{equation}

The spectrum of the glottal source open phase is then computed from zeros outside the unit circle (anticausal component) while zeros inside it give the vocal tract transmittance modulated by the source return phase spectrum (causal component). To obtain such a separation, the effects of the windowing are known to play a crucial role \cite{windowing}. In particular, we have shown that a Blackman window centered on the Glottal Closure Instant (GCI) and whose length is twice the pitch period is appropriate in order to achieve a good decomposition.

\section{Chirp Decomposition of Speech}\label{sec:ZCZT}

The Chirp Z-Transform (CZT), as introduced by Rabiner et al \cite{chirp} in 1969, allows the evaluation of the Z-transform on a spiral contour in the Z-plane. Its first application aimed at separating too close formants by reducing their bandwidth. Nowadays CZT reaches several fields of Signal Processing such as time interpolation, homomorphic filtering, pole enhancement, narrow-band analysis,...

As previously mentioned, the ZZT-based decomposition is strongly dependent on the applied windowing. This sensitivity may be explained by the fact that ZZT implicitly conveys phase information, for which time alignment is known to be crucial \cite{Tribolet}. In that article, it is observed that the window shape and onset may lead to zeros whose topology can be detrimental for accurate pulse estimation. The subject of this work is precisely to handle with these zeros close to the unit circle, such that the ZZT-based technique correctly separates the causal (i.e minimum-phase) and anticausal (i.e maximum-phase) components.
 
\begin{figure}[!ht]
  % Requires \usepackage{graphicx}
  \centering
  \includegraphics[width=1\textwidth]{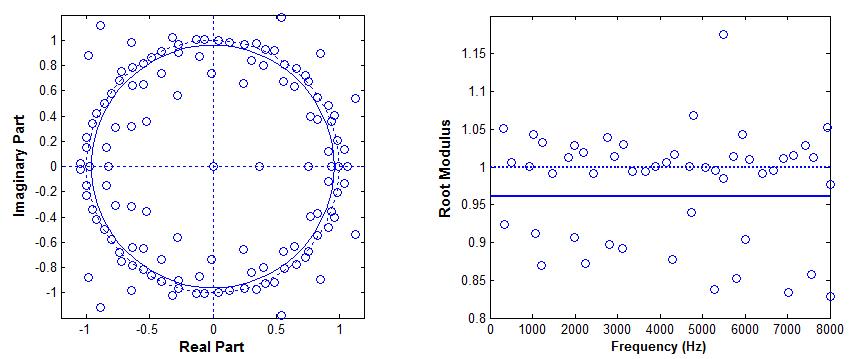}
  \caption{Example of root distribution for a natural speech frame. \emph{Left panel}: representation in the Z-plane, \emph{Right panel}: representation in polar coordinates. The chirp circle (solid line) allows a correct decomposition, contrarily to unit circle (dotted line).}
  \label{fig:Distri}
\end{figure}

For this, we evaluate the CZT on a circle whose radius $R$ is chosen so as to split the root distribution into two well-separated groups. More precisely, it is observed that the significant impulse present in the excitation at the GCI results in a gap in the root distribution. When analysis is exactly GCI-synchronous, the unit circle perfectly separates causal and anticausal roots. On the opposite, when the window moves off from the GCI, the root distribution is transformed. Such a decomposition is then not guaranteed for the unit circle and another boundary is generally required. Figure \ref{fig:Distri} gives an example of root distribution for a natural voiced speech frame for which an error of 0.6 ms is made on the real GCI position. It is clearly seen that using the traditional ZZT-based decomposition ($R=1$) for this frame will lead to erroneous results. In contrast, it is possible to find an optimal radius leading to a correct separation. 

%As a means for automatically determining such a radius, the sorted root moduli are inspected and the greatest discontinuity in the unit circle
% vicinity is detected. The chosen radius $R$ is then chosen as the middle of this discontinuity, and is assumed to give the best separation 
% between the causal and anticausal roots.

In order to automatically determine such a radius, let us have the following thought process. We know that ideally the analysis should be GCI-synchronous. When this is not the case, the chirp analysis tends to modify the window such that its center coincides with the nearest GCI (to ensure a reliable phase information). Indeed, evaluating the chirp Z-transform of a signal $x(t)$ on a circle of radius $R$ is equivalent to evaluating the Z-transform of $x(t)\cdot{}exp(log(1/R)\cdot{}t)$ on the unit circle. It can be demonstrated that for a Blackman window $w(t)$ of length $L$ starting in $t=0$:

\begin{equation}\label{eq:Black}
w(t)=0.42-0.5\cdot{}\cos(\frac{2\pi t}{L})+0.08\cdot{}\cos(\frac{4\pi t}{L}),
\end{equation}

the radius $R$ necessary to modify its shape so that its new maximum lies in position $t^{*}$ ($<L$) is expressed as:

\begin{equation}\label{eq:IdealZCZT}
R=exp[\frac{2\pi}{L}\cdot{}\frac{41\tan^2(\frac{\pi t^{*}}{L})+9}{25\tan^3(\frac{\pi t^{*}}{L})+9\tan(\frac{\pi t^{*}}{L}))}].
\end{equation}

In particular, we verify that $R=1$ is optimal when the window is GCI-centered ($t^{*}=\frac{L}{2}$) and, since we are working with two-period long windows, the optimal radius does not exceed $exp(\pm\frac{50\pi}{17L})$ in the worst cases (the nearest GCI is then positioned in $t^{*}=\frac{L}{4}$ or $t^{*}=\frac{3L}{4}$). As a means for automatically determining the radius allowing an efficient separation, the sorted root moduli are inspected and the greatest discontinuity in the interval $[exp(-\frac{50\pi}{17L}),exp(\frac{50\pi}{17L})]$ is detected. Radius $R$ is then chosen as the middle of this discontinuity, and is assumed to optimally split the roots into minimum and maximum-phase contributions.

\begin{figure}[!ht]
  % Requires \usepackage{graphicx}
  \centering
  \includegraphics[width=1\textwidth]{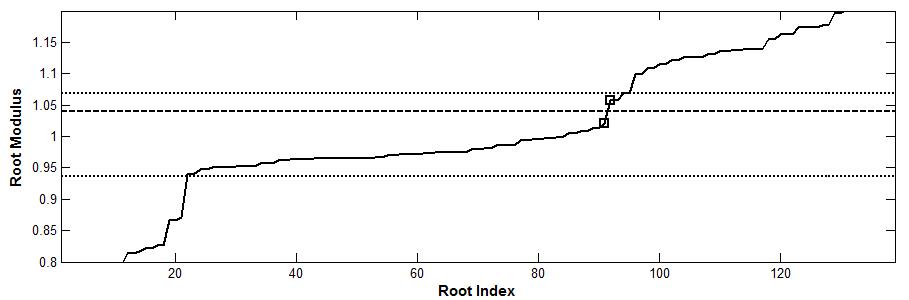}
  \caption{Determination of radius $R$ (dashed line) for ZCZT computation by detecting, within the bounds $exp(\pm\frac{50\pi}{17L})$ (dotted lines), a discontinuity (indicated by rectangles) in the sorted root moduli (solid line).}
  \label{fig:Sorted}
\end{figure}

\section{Experimental Results}\label{sec:results}

This Section gives a comparative evaluation of the following methods:
 
\begin{itemize}
\item \emph{the traditional ZZT-based technique}: $R=1$,
\item \emph{the proposed ZCZT-based technique}: $R$ is computed as explained at the end of Section \ref{sec:ZCZT} (see Fig. \ref{fig:Sorted}),
\item \emph{the ideal ZCZT-based technique}: $R$ is computed from Equation \ref{eq:IdealZCZT} where the real GCI location $t^{*}$ is known. This can be seen as the ultimate performance one can expect from the ZCZT-based technique.
\end{itemize} 
 
Among others, it is emphasized how the proposed technique is advantageous in case of GCI location errors.
 
\subsection{Results on Synthetic Speech}\label{ssec:Synth}

Objectively and quantitatively assessing a method of glottal signal estimation requires working with synthetic signals, since the real source is not available for real speech signals. In this work, synthetic speech signals are generated for different test conditions, by passing a train of Liljencrants-Fant waves \cite{LF} through an all-pole filter. This latter is obtained by LPC analysis on real sustained vowel uttered by a male speaker. In order to cover as much as possible the diversity one can find in real speech, parameters are varied over their whole range. Table \ref{tab:ParamRange} summarizes the experimental setup. Note that since the mean pitch during the utterances for which the LP coefficients were extracted was about 100 Hz, it reasonable to consider that the fundamental frequency should not exceed 60 and 180 Hz in continuous speech.

\begin{table}[!ht]
\centering
\begin{tabular}{|c|c|c|c|c|}
   \hline
   \multicolumn{3}{|c|}{Source Characteristics} & Filter & Perturbation \\       
   \hline
   Open Quotient & Asymmetry Coeff. & Pitch & Vowel & GCI location error \\
   \hline
   0.4:0.05:0.9 & 0.6:0.05:0.9 & 60:20:180 Hz & /a/,/e/,/i/,/u/ & -50:5:50 \% of $T_0$ \\
   \hline   
\end{tabular}
\newline
\caption{Details of the test conditions for the experiments on synthetic signals.}
\label{tab:ParamRange}
\end{table}

To evaluate the performance of our methods, two objective measures are used:

\begin{itemize}
\item \emph{the determination rate on the glottal formant frequency $F_g$}: As one of the main feature of the glottal source, the glottal formant \cite{Doval} should be preserved after estimation. The determination rate consists of the percentage of frames for which the relative error made on $F_g$ is lower than 20\%.
\item \emph{the spectral distortion}: This measure quantifies in the frequency-domain the distance between the reference and estimated glottal waves (here noted $x$ and $y$ by simplification), expressed as:

\begin{equation}\label{eq:SSD}
SD(x,y) = \sqrt{\int_{-\pi}^\pi(20\log_{10}|\frac{X(\omega)}{Y(\omega)}|)^2\frac{\emph{d}\omega}{2\pi}}
\end{equation}

\end{itemize}
\begin{figure}[!ht]
  % Requires \usepackage{graphicx}
  \centering
  \includegraphics[width=0.9\textwidth]{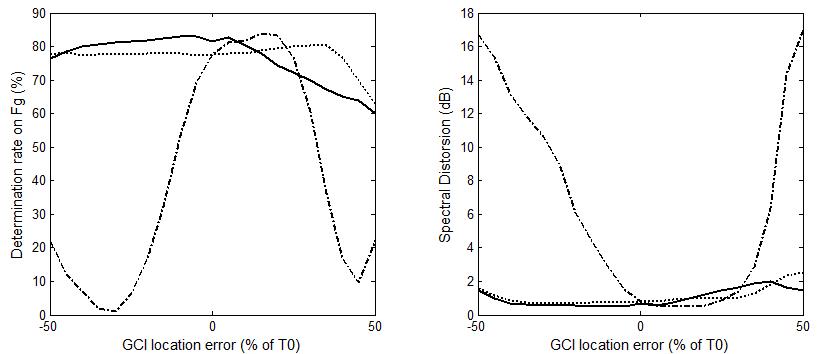}
  %\centering
  \caption{Comparison of the traditional ZZT (dashdotted line), proposed ZCZT (solid line) and ideal ZCZT (dotted line) based methods on synthetic signals according to their sensitivity to an error on the GCI location. \emph{Left panel:} Influence on the determination rate on the glottal formant frequency. \emph{Right panel:} Influence on the spectral distortion.}
  \label{fig:Synth}
\end{figure}

Figure \ref{fig:Synth} compares the results obtained for the three methods according to their sensitivity to the GCI location. The proposed ZCZT-based technique is clearly seen as an enhancement of the traditional ZZT approach when an error on the exact GCI position is made.

\subsection{Results on Real Speech}\label{ssec:Real}

\begin{figure}[!ht]
  % Requires \usepackage{graphicx}
  \centering
  \includegraphics[width=1\textwidth]{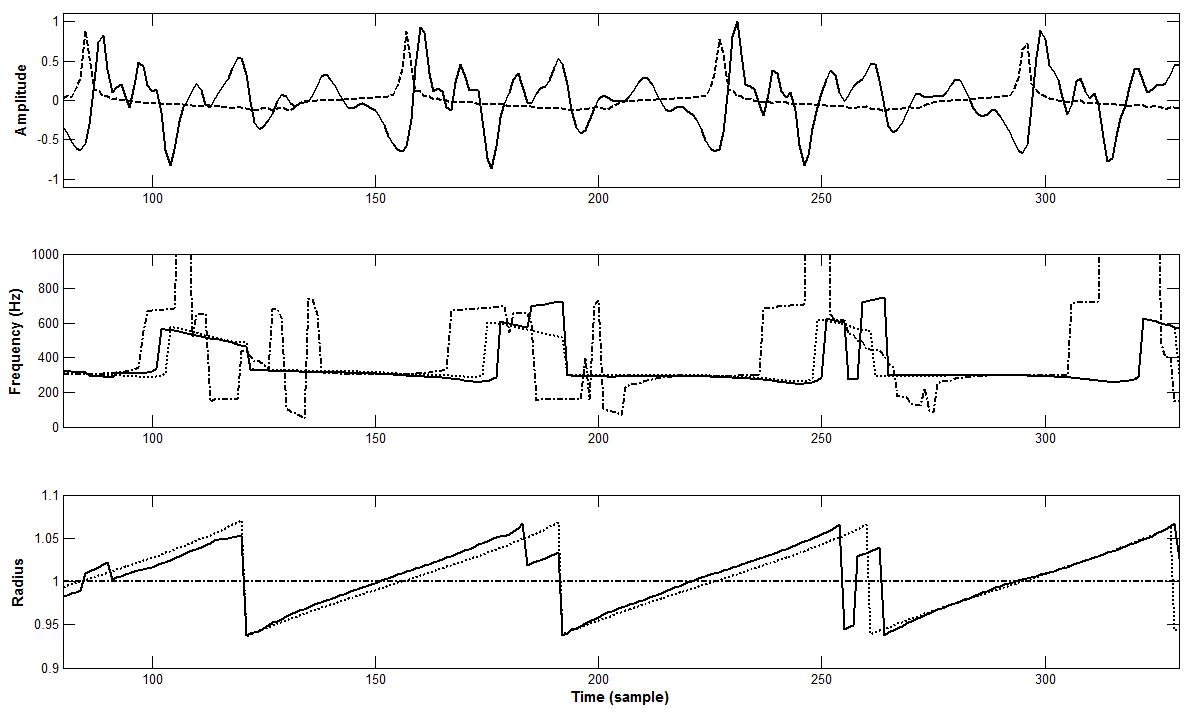}
  \caption{Comparison of ZZT and ZCZT-based methods on a real voiced speech segment. \emph{Top panel:} the speech signal (solid line) with the synchronized differenced EGG (dashed line). \emph{Middle panel:} the glottal formant frequency estimated by the traditional ZZT (dashdotted line), the proposed ZCZT (solid line) and the ideal ZCZT (dotted line) based techniques. \emph{Bottom panel:} Their corresponding radius used to compute the chirp Z-transform.}
  \label{fig:Real}
\end{figure}

\begin{figure}[!ht]
  % Requires \usepackage{graphicx}
  \centering
  \includegraphics[width=1\textwidth]{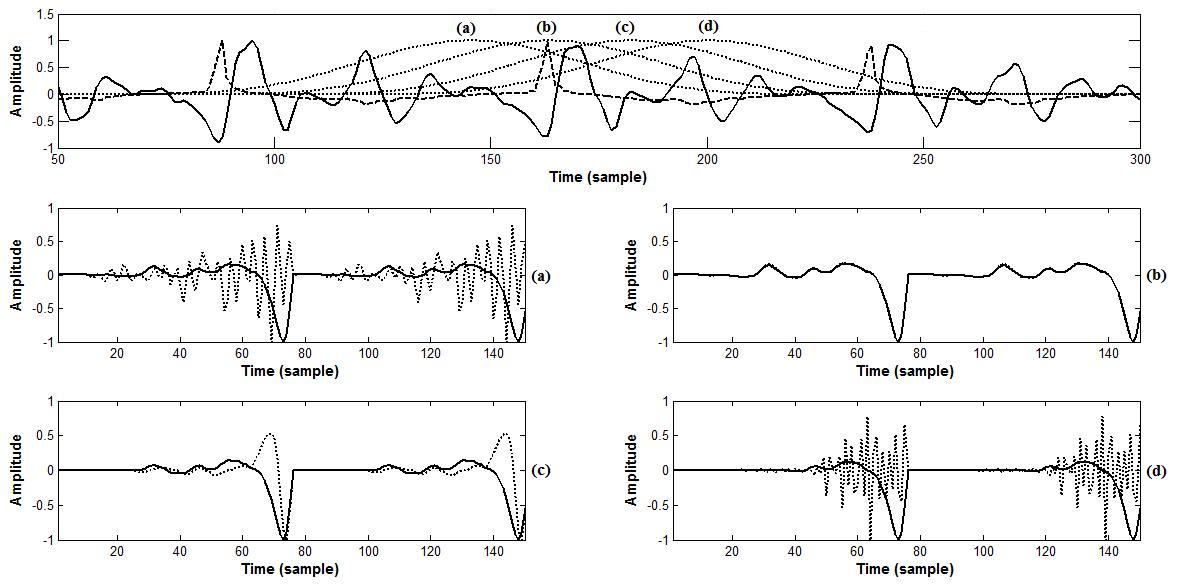}
  \caption{Examples of glottal source estimation using either the traditional ZZT or the proposed ZCZT-based method. \emph{Top panel:} a voiced speech segment (solid line) with the synchronized differenced EGG (dashed line) and four different positions of the window (dotted line). \emph{Panels (a) to (d):} for the corresponding window location, two cycles of the glottal source estimation achieved by the traditional ZZT (dotted line) and by the proposed ZCZT-based technique (solid line).}
  \label{fig:4Decomp}
\end{figure}

Figure \ref{fig:Real} displays an example of decomposition on a real voiced speech segment (vowel /e/ from \emph{BrianLou4.wav} of the Voqual03 database, $F_s=16kHz$). The top panel exhibits the speech waveform together with the synchronized (compensation of the delay between the laryngograph and the microphone) differenced Electroglottograph (EGG) informative about the GCI positions. Both next panels compare respectively the detected glottal formant frequency $F_g$ and the radius for the three techniques. In the middle panel, deviations from the constant $F_g$ can be considered as errors since $F_g$ is expected to be almost constant during three pitch periods. It may be noticed that the traditional ZZT-based method degrades if analysis is not achieved in the GCI close vicinity. Contrarily, the proposed ZCZT-based technique gives a reliable estimation of the glottal source on a large segment around the GCI. Besides the obtained performance is comparable to what is carried out by the ideal ZCZT. In Figure \ref{fig:4Decomp} the glottal source estimated by the traditional ZZT and the proposed ZCZT-based method are displayed for four different positions of the window (for the vowel /a/ from the same file). It can be observed that the proposed technique (solid line) gives a reliable estimation of the glottal flow wherever the window is located. On the contrary the sensivity of the traditional approach can be clearly noticed since its glottal source estimation turns out to be irrelevant when the analysis is not performed in a GCI-synchronous way.

\section{Conclusion}\label{sec:conclu}
This paper proposed an extension of the ZZT-based technique we proposed in \cite{ZZT}. The enhancement consists in evaluating the Z-transform on a contour possibly different from the unit circle. For this we considered, in the Z-plane, circles whose radius is automatically determined by detecting a discontinuity in the root distribution. It is expected that such circles lead to a better separation of both causal and anticausal contributions. Results obtained on synthetic and real speech signals report an advantage for the proposed ZCZT-based technique, mainly when GCIs are not accurately localized. As future work, we plan to characterize the glottal source based on the proposed framework.

\section*{Acknowledgments}\label{sec:Acknowledgments}

Thomas Drugman is supported by the ``Fonds National de la Recherche Scientifique'' (FNRS).


\begin{thebibliography}{1}

\bibitem {Speaker}
M. Plumpe, T. Quatieri, D. Reynolds: \emph{Modeling of the glottal flow derivative waveform with application to speaker identification}
IEEE Trans. on Speech and Audio Processing, vol. 7, pp. 569–586, 1999.

\bibitem {Disorder}
E. Moore, M. Clements, J. Peifer, L. Weisser: \emph{Investigating the role of glottal features in classifying clinical depression}, Proc. of the 25th International Conference of the IEEE Engineering in Medicine and Biology Society, vol. 3, pp. 2849-2852, 2003

\bibitem {Reco}
D. Yamada, N. Kitaoka, S. Nakagawa: \emph{Speech Recognition Using Features Based on Glottal Sound Source}, Trans. of the Institute of Electrical Engineers of Japan, vol. 122-C, no. 12, pp. 2028-2034, 2002.

\bibitem {Synth}
T. Drugman, G. Wilfart, A. Moinet, T. Dutoit: \emph{ Using a pitch-synchronous residual for hybrid HMM/frame selection speech synthesis}, Proc. IEEE International Conference on Speech and Signal Processing, 2009.

\bibitem {DAP}
P. Alku, E. Vilkman: \emph{Estimation of the glottal pulseform based on discrete all-pole modeling}, Third International Conference on Spoken Language Processing, pp. 1619-1622, 1994.

\bibitem {IAIF}
P. Alku, J. Svec, E. Vilkman, F. Sram: \emph{Glottal wave analysis with pitch synchronous iterative adaptive
inverse filtering}, Speech Communication, vol. 11, issue 2-3, pp. 109-118, 1992.


\bibitem {ClosedPhase}
D. Veeneman, S. BeMent: \emph{Automatic glottal inverse filtering from
speech and electroglottographic signals}, IEEE Trans. on Signal Processing, vol. 33, pp. 369–377, 1985.

\bibitem {MCLPC}
D. Brookes, D. Chan: \emph{Speaker characteristics from a glottal airflow model using glottal inverse filtering}, Proc. Institue of Acoust., vol. 15, pp. 501–508, 1994.

\bibitem {ZZT}
B. Bozkurt, B. Doval, C. D'Alessandro, T. Dutoit: \emph{Zeros of Z-Transform Representation With Application to Source-Filter Separation in Speech}
IEEE Signal Processing Letters, vol. 12, no. 4, 2005.

\bibitem {Doval}
B. Doval, C. d'Alessandro, N. Henrich: \emph{The voice source as a causal/anticausal linear filter}, Proceedings ISCA ITRW VOQUAL03, pp. 15-19, 2003.

\bibitem {Sturmel}
N. Sturmel, C. D'Alessandro, B. Doval: \emph{A comparative evaluation of the Zeros of Z-transform representation for voice source estimation}, The Interspeech07, pp. 558-561, 2007.

\bibitem {windowing}
B. Bozkurt, B. Doval, C. D'Alessandro, T. Dutoit: \emph{Appropriate windowing for group delay analysis and roots of Z-transform of speech signals}, Proc. of the 12th European Signal Processing Conference, 2004.

\bibitem {chirp}
L. Rabiner, R. Schafer, C. Rader: \emph{The Chirp-Z transform Algorithm and Its Application}, Bell System Technical Journal, vol. 48, no.5, pp. 1249-1292, 1969.

\bibitem {Tribolet}
J. Tribolet, T. Quatieri, A. Oppenheim: \emph{Short-time homomorphic analysis}, IEEE International Conference on Speech and Signal Processing, vol. 2, pp.716-722, 1977. 

\bibitem {LF}
G. Fant, J. Liljencrants, Q. Lin: \emph{A four parameter model of glottal flow}, STL-QPSR4, pp. 1-13, 1985.

\end{thebibliography}
\end{document}